# Investigation of the Dzyaloshinskii-Moriya interaction and room temperature skyrmions in W/CoFeB/MgO thin films and microwires

S. Jaiswal[1,2], K. Litzius[1,3,4], I. Lemesh[5], F. Büttner[5], S. Finizio[6], J. Raabe[6], M. Weigand[7],
K. Lee[1], J. Langer[2], B. Ocker[2], G. Jakob[1,3], G. S. D. Beach[5], M Kläui[1,3]

[1]Johannes Gutenberg Universität-Mainz, Institut für Physik, Staudinger Weg 7, 55128 Mainz, Germany
[2]Singulus Technologies AG, 63796 Kahl am Main, Germany
[3]Graduate School of Excellence "Materials Science in Mainz" (Mainz), Staudinger Weg 9, 55128 Mainz, Germany
[4]Max Plank Institute for intelligent systems, 70569 Stuttgart, Germany
[5]Department of Materials Science and Engineering, Massachusetts Institute of Technology, Cambridge, Massachusetts 02139, USA
[6]Swiss Light Source, Paul Scherrer Institut, Villigen PSI CH-5232, Switzerland
[7]Max Planck Institute for Intelligent Systems, 70569 Stuttgart, Germany

Recent studies have shown that material structures, which lack structural inversion symmetry and have high spin-orbit coupling can exhibit chiral magnetic textures and skyrmions which could be a key component for next generation storage devices. The Dzyaloshinskii-Moriya Interaction (DMI) that stabilizes skyrmions is an anti-symmetric exchange interaction favoring non-collinear orientation of neighboring spins. It has been shown that material systems with high DMI can lead to very efficient domain wall and skyrmion motion by spin-orbit torques. To engineer such devices, it is important to quantify the DMI for a given material system. Here we extract the DMI at the Heavy Metal (HM) /Ferromagnet (FM) interface using two complementary measurement schemes namely asymmetric domain wall motion and the magnetic stripe annihilation. By using the two different measurement schemes, we find for W(5 nm)/$Co_{20}Fe_{60}B_{20}$(0.6 nm)/MgO(2 nm) the DMI to be 0.68 ± 0.05 mJ/m$^2$ and 0.73 ± 0.5 mJ/m$^2$, respectively. Furthermore, we show that this DMI stabilizes skyrmions at room temperature and that there is a strong dependence of the DMI on the relative composition of the CoFeB alloy. Finally we optimize the layers and the interfaces using different growth conditions and demonstrate that a higher deposition rate leads to a more uniform film with reduced pinning and skyrmions that can be manipulated by Spin-Orbit Torques.

Recent advances in thin film fabrication processes have led to the accelerated development of magnetic storage devices. This has opened exciting areas of research due to the effects occurring at the interface between a heavy metal (HM) and a ferromagnet (FM). This interface is the building block for next generation memory devices such as the Spin-Orbit Torque (SOT) MRAM[1–4]. There are a number of important phenomena associated with the interface[5]: interfacial contributions to the SOTs[6], interfacial perpendicular anisotropy[7,8], and interfacial Dzyaloshinskii-Moriya interaction (DMI)[9–12]. DMI is an anti-symmetric exchange interaction which favours non-collinear alignment of neighbouring spins $S_1$ and $S_2$, whose magnitude is defined by the DMI vector **D**. This anti-symmetric exchange interaction favours chiral canting of spins which lead to special chiral spin textures[13] and in particular magnetic skyrmions[14–16].

Recent studies have demonstrated that Néel-like skyrmions are stabilized in thin films possessing an interfacial DMI where the symmetry breaks at the interface between the HM and the FM[17–21]. Such skyrmions have been envisaged to be used in skyrmion based racetrack memory[22,23] due to their topologically enhanced stability and low threshold current densities for propagation[18,22,24,25]. These low current densities however, have only been found for motion of skyrmion lattices[26]. While conventional spin transfer torque effects may also occur in such stacks the contribution is not sizeable[27]. For memory devices, it is imperative to achieve meta-stable skyrmions at room temperature. Only those allow for writing and deleting processes, such that both the skyrmion and the single domain state are stable in materials compatible with CMOS technology[4]. So far only few systems have been identified to host such skyrmions. However, these systems are plagued by a large number of pinning sites which prevent the study of skyrmion dynamics. Given the importance of DMI for memory applications[28,29], it is essential that it is quantified in different material systems using reliable techniques. Earlier works have used current induced DW motion (CIDWM) to estimate the DMI in thin film microstructures[30]. However, CIDWM has different components of current dependent spin-torques associated with it that can all move DWs and skyrmions making the analysis not straight forward[16].

In this paper, we address both the thin film deposition and the quantification of DMI in W/CoFeB/MgO layers. We demonstrate that identical thicknesses of films, grown with different conditions, may yield different values for the DMI. This is related to the underlying crystal structure that can be modified by tuning the film deposition conditions. We show that this material system exhibits room temperature skyrmions and it is used to systematically study the DMI. We use two different methods to evaluate the DMI in the same material stacks, namely the asymmetric field driven domain expansion in thin films[31,32] and the magnetic domain stripe annihilation[20,21] in nanostructures. Finally we use the optimal stack and demonstrate the presence of metastable skyrmions, which are moved by current pulses.

The studied samples were substrate/W(5 nm)/$Co_{20}Fe_{60}B_{20}$(0.6 nm)/MgO(2 nm) continuous thin films grown on thermally oxidised silicon substrates. All thin film materials were sputter deposited using a *Singulus Rotaris* deposition system with a base pressure < 3 x$10^{-8}$ mbar. A 5 nm Ta cap was used to prevent oxidation of the film. Additionally continuous multilayer films of substrate/[W(5 nm)/$Co_{20}Fe_{60}B_{20}$(0.6 nm)/MgO(2 nm)]$_{15}$/Ta(5) were grown on SiN membranes and patterned into microwires of 1.4 µm x 5.0 µm using e-beam lithography followed by lift-off. Kerr microscopy was used to observe the magnetic domain expansion on the single layered thin film, and Scanning Transmission X-ray Microscopy (STXM) with X-ray Magnetic Circular Dichroism (XMCD) contrast was employed to study the room temperature skyrmions in the microwires.

In order to investigate the field driven motion of magnetic domains reliably, it is crucial to obtain an out-of-plane (OOP) domain nucleation with a minimum number of point defects within the material stack which may lead to pinning of the domain walls during field propagation[33].

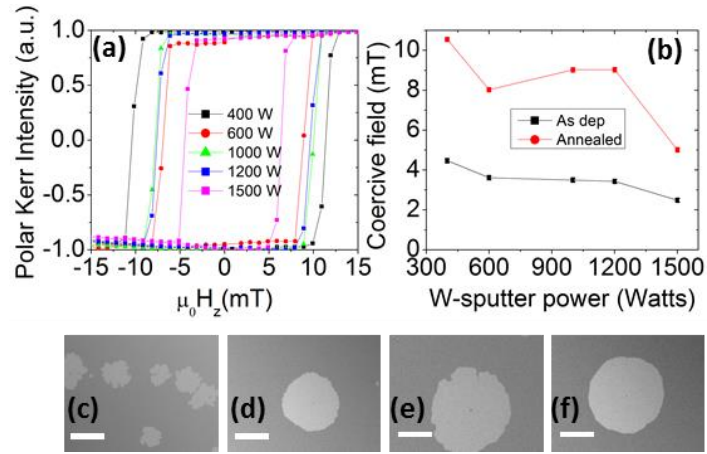

Fig. 1. OOP magnetisation curves (a) for annealed samples with different W sputter powers for $t_{FM}$ = 0.6 nm. Coercivities (b) for both as deposited and annealed thin films for $t_{FM}$ = 0.6 nm. Error bars are smaller than the symbols. Differential Kerr microscopy images of as deposited films with W sputtered at (c) 400 W, (e) 1500 W and annealed films of the same material stack (d) 400 W, (f) 1500 W. The scale bar represents 100 μm.

Therefore, the material stack was optimised by growing the ferromagnet on different W layers for which the W sputter power was varied while keeping the thickness constant. This allowed for the structural modification of the HM/FM interface. Additionally, the films were annealed at 400° C for 2 h in vacuum. As shown in Fig. 1(a) all films show an easy axis orientation of the magnetization in the OOP direction as measured by polar MOKE magnetometry. The coercivity, shown in Fig. 1(b) for the annealed samples, is at least twice the value of the as-deposited samples with the exception of the sample sputtered at 1500 W. This suggests a possible transformation of the CoFeB from the amorphous phase into the polycrystalline phase during annealing[34–36]. Also the coercive field decreases for higher sputter power. Using a higher sputtering power density for the growth of the seed layer leads to a greater density of target atoms in the plasma. This may lead to a higher density of nucleation sites due to a higher supersaturation of the target atom species[37] and thereby, facilitating a smoother growth of the respective layer. The domain structures for the as-deposited and annealed states are shown in Fig. 1(c-f). Multiple nucleation points are observed in the 400 W sputtered thin film due to defects in the growth of the thin film. The domain structure smoothens, indicating a reduction in pinning as the sputter power is increased to 1500 W and finally lowest pinning is attained once the samples are annealed. For a given sputter power (Fig. 1(c-f)) annealing also induces a smoothening of the domain structures. Therefore, annealing here leads to a domain configuration with reduced pinning, which facilitates field driven experiments to measure the DMI. Recent studies[38] on a similar materials stack suggest that the DMI decreases with increasing annealing temperatures. Therefore, while annealing is crucial to obtain a reduced pinning, it may also decrease the DMI value in such ferromagnetic alloys based material stacks.

Prior to measurements, the sample is saturated in the OOP direction and a bubble domain is nucleated by applying an OOP field in the opposite direction (Fig 2a). A static in-plane field is then applied which leads to the asymmetry in the DW motion.

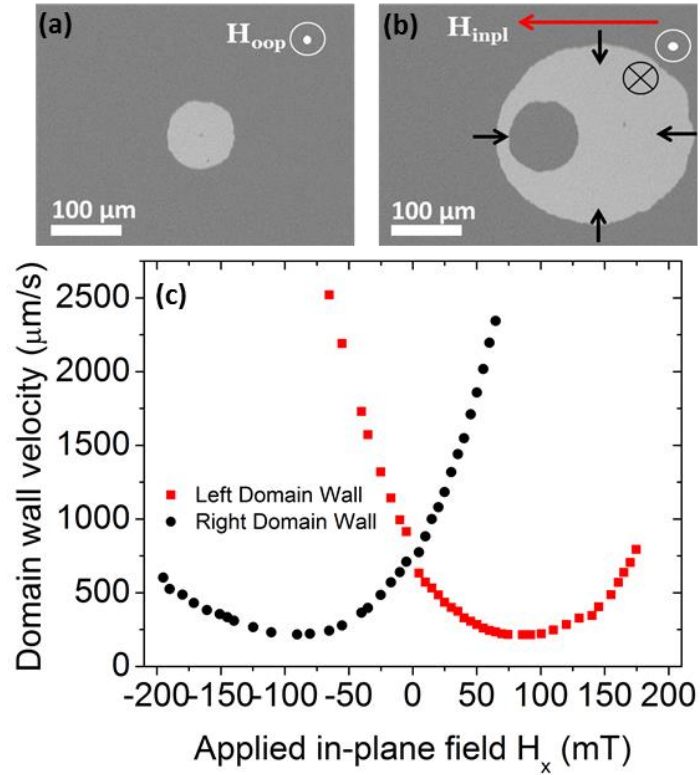

Fig. 2. Differential Kerr microscopy images for (a) isometric expansion with only an OOP field and (b) asymmetric expansion with both OOP and in-plane field. (c) Domain wall velocities for both left and right moving domain walls.

The images before and after the application of the in-plane field are subtracted which results in a difference image shown in Fig. 2b. The region along the centre of the OOP nucleated domain is studied and the velocities are calculated by measuring the domain wall displacement and the known pulse duration. This procedure was repeated for each in-plane field value at least four times. In the absence of an in-plane field, the domain wall maintains a radial symmetry. However, when an in-plane field is applied, the symmetry is broken and the domain walls moving parallel and anti-parallel to the in-plane field exhibit different velocities due to the DMI effective field[32] (Fig. 2c). The DMI effective field assists the motion of one wall while hindering the other. The in-plane field at which the domain walls experience a minimum velocity is the effective DMI field. Note that the velocity at this field value is not zero as there is still an OOP field being applied. Each domain wall velocity minimum is offset from zero and this offset is taken as the DMI field. The DMI constant $D$ is directly dependent on the effective DMI field as, $\mu_o H_{DMI} = D/M_s \Delta$, where $\Delta$ is the domain wall width (~6 nm). The saturation magnetization $M_s = (1.14 \pm 0.04) \cdot 10^6$ A/m and anisotropy field $\mu_0 H_K = 400$ mT were measured using vibrating sample magnetometry. The domain wall width is defined as $\sqrt{A/K_{eff}}$ where $K_{eff}$ is the effective anisotropy of the perpendicularly magnetized system and is defined as $K_{eff} = (\mu_0 M_s H_K)/2$ which takes into account the demagnetising field and assuming $A$ the exchange stiffness being 10 pJ/m for the ferromagnet used. The resulting DMI constant is $D = (0.68 \pm 0.05)$ mJ/m$^2$ with an effective DMI field $93.0 \pm 0.1$ mT. Compared to nominally similar material stacks that have been investigated by Soucaille *et al.*[39], our value for the DMI constant for W/CoFeB films is twice of what they obtained. However, it is important to note that the DMI scales on the atomic level and even slight differences in interface quality can produce a dramatic difference in the resulting value. Moreover, their FM is nominally 1.7 times thicker which may explain the lower DMI value they obtained. To probe the influence of the FM composition, the DMI was measured also for a nominally similar film but an alloy with an equal amount of Co and Fe, namely W(5 nm)/Co$_{40}$Fe$_{40}$B$_{20}$(0.6 nm)/MgO(2 nm)/Ta(5 nm). The DMI was calculated using the asymmetric domain expansion and found to be $0.028 \pm 0.05$ mJ/m$^2$, with an effective DMI field $4.0 \pm 0.1$ mT, thereby indicating a strong influence of the ferromagnetic alloy composition on the DMI. Such a compositional dependence of DMI has not yet been reported, but this highlights the subtle effect that governs the DMI.

In order to obtain magnetic domain information at smaller length and time scales and to achieve a complementary value for the DMI, we performed STXM experiments at the synchrotron while exploiting the XMCD for element selective magnetic contrast. Fig. 3a shows STXM with XMCD images of stripe domain structures in multilayers of subs/[W(5 nm)/Co$_{20}$Fe$_{60}$B$_{20}$(0.6 nm)/MgO(2 nm)]$_{15}$/Ta(5 nm) with an effective ferromagnetic thickness of 9 nm. The X-ray energy was tuned to the Fe absorption edge and the measurement was performed in a perpendicular geometry.

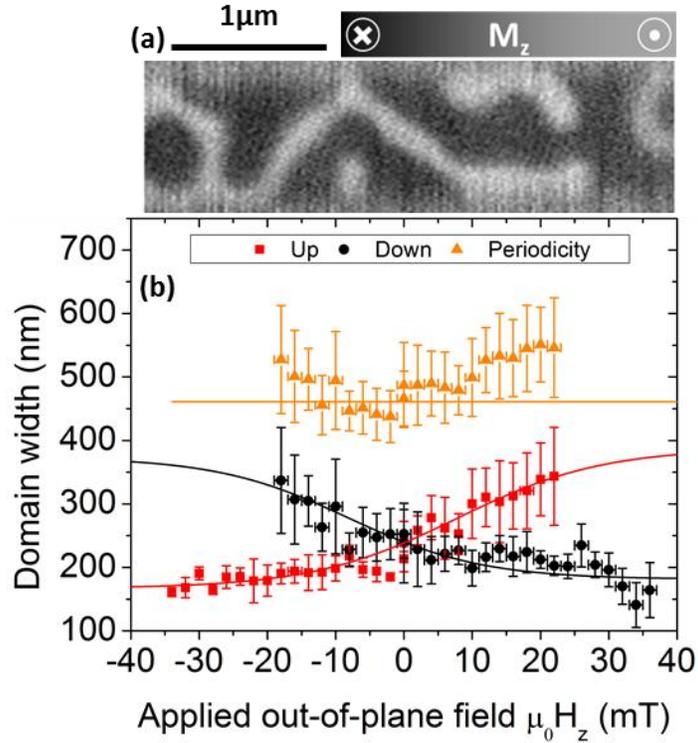

Fig. 3. (a) STXM with XMCD image of magnetic stripe domains of the sample [W(5 nm)/CoFeB(0.6 nm)/MgO(2 nm)]$_{15}$/Ta(5 nm) in a 1.4 µm x 5.0 µm micro-wire.(b) Stripe domain width (up and down domains) variation as a function of applied field.

The magnetisation was initially saturated in the OOP direction by applying a field of 50 mT and then the applied field was gradually decreased until a negative field resulted in worm like domains shown in Fig. 3a. The magnetic domain stripe width was measured as a function of OOP applied field; this field dependence is shown in Fig. 3b. It can be seen from Fig. 3b that as the field is increased further in the negative direction the stripe widths of up polarity (red curve) decrease and gradually the up polarity stripe domains annihilate as the sample attains saturation. The domain width at the maximum field value at which the worm domains still exist is used to determine the DMI[20]. By investigating the field evolution of the domain width one obtains a hysteresis loop (Fig. 3b) which can be fitted using the function, $w(H) = a \cdot \tanh(\delta \cdot H + \phi) + d$. Here, $w(H)$ is the domain width as a function of external applied field, $\delta$ is the inverse loop width, $\phi$ the phase offset, $a$ the amplitude, and $d$ the domain width at zero field[21].

As the field approaches the saturation region, the magnetic stripe domains of opposite polarity (black and white contrast) corresponding to up and down magnetization orientation approach a terminal width before annihilation of the domain wall. The terminal widths are extracted from $w_{term} = |d - a|$ and result in values of $w_{term}$ = 172 ± 24 nm and an average periodicity of $w_{aver}$ = 461 ± 10 nm. Once the terminal width of the stripe expansion is calculated, the value of *D* can be computed by minimising the total effective energy density of the multilayer film,

$$\mathcal{E}_{tot,eff}^{\infty,N} = \frac{1}{w}\left[\frac{2A'}{\Delta} + 2K'_u\Delta - \pi D'\right] + C + \mathcal{E}_{d,s} + \mathcal{E}_{d,v} \quad (1)$$

with the surface and the volume stray field energies defined as:

$$\mathcal{E}_{d,s} = \chi \sum_{n=1,3,5..}^{\infty} \frac{1}{\left[\sinh\left(\frac{\pi^2 n\Delta}{2w}\right)\right]^2} \frac{1-e^{-\upsilon}}{n} \quad (2)$$

$$\mathcal{E}_{d,v} = \chi \sum_{n=1,3,5..}^{\infty} \frac{1}{\left[\cosh\left(\frac{\pi^2 n\Delta}{2w}\right)\right]^2} \frac{e^{-\upsilon}+\upsilon-1}{n} \quad (3)$$

and the effective constants are defined as: $A' = fA$, $D' = fD$, $M'_s = fM_s$, $K'_u = K_u f - \frac{\mu_o M_s^2}{2}(f-f^2)$, $C = \frac{\mu_o M_s^2}{2}(f-f^2)$, the constants in eqs. (2-3) are defined as $\chi = \frac{\pi\mu_o M_s^2 \Delta^2}{w\lambda}$ and $\upsilon = \pi n\lambda/w$ where $f$ is the scaling factor given by the ratio of thickness of a single FM layer thickness to the multilayer periodicity, $\lambda$ is the product of multilayer periodicity and the number of layer repeats and $K_u$ is the uniaxial anisotropy. The surface and volume contributions due to the use of a multi-layered sample are taken into account in eq. (2) and eq. (3). The theoretical model of the stripe domain phase used here is described in[40]. The value of *D* as computed from the stripe annihilation method was $0.73 \pm 0.5$ mJ/m$^2$, which is in line with the value determined using the asymmetric domain expansion. The relative errors in the DMI values using the two methods described are different. This is because for the bubble expansion measurements, magnetic domains which showed no significant pinning but smooth elliptic expansion were used. The total error is governed by the minimum velocity determination. However for wires all stripe domains were evaluated including regions with pinning sites which dominates the DMI error. It has been shown for systems with such large DMI the domain walls have a full Néel character. However, given the small domain wall widths it is not possible to resolve it using X-ray microscopy techniques. Having determined a significant DMI, the next step is to study skyrmions in this system. In order to nucleate skyrmions from the stripe domain phase of the system and move them, we applied current pulses with a length of ~5 ns and a current density of $3.9 \cdot 10^{11}$ Am$^{-2}$. This resulted in the transformation of the stripe domains into Néel skyrmions in the magnetic wires as previously shown[20].

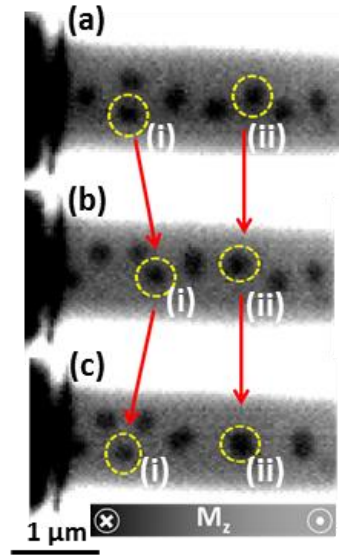

Fig. 4. Room temperature skyrmions nucleated in 1.4 µm x 5.0 µm micro-wire of [W(5 nm)/CoFeB(0.8 nm)/MgO(2 nm)]$_{10}$/Ta(5 nm) (a). Subsequently nucleated skyrmions are moved forwards and backwards by applying current pulses with opposite polarity (b)-(c). Highlighted skyrmion (i) is moved forward (current applied along the wire from left to right) and backward (reversed polarity). Skyrmion (ii) is an example of a pinned skyrmion that does not move at this current density.

In Fig. 4a we present room temperature skyrmions nucleated in the W/CoFeB/MgO thin films wire. Current pulses were applied along the length of the wire in order to move the skyrmions. We show a selected example in Fig. 4b-c. The pulse direction was reversed and we find as one example that the skyrmion (i) moves back and forth depending on the direction of the applied current pulse as shown in Fig. 4b-4c, while as another example skyrmion (ii) remains pinned in the wire and does not move irrespective of the current pulse. The direction of motion with respect to the current direction indicates here a right handed DMI (*D*>0) (see supplementary material in Ref[20]). The skyrmions are observed to move against electron flow. and show generally a regular circular shape in line with the low pinning deduced from our domain wall motion experiments.

In conclusion, we have developed a perpendicular magnetization multilayer system which exhibits room temperature skyrmions. We have shown that tuning the deposition conditions of the W seed layer and annealing of the material stack allows for a systematic improvement of the smoothness of the domain structures and thus a reduction of the pinning. The DMI for this material stack was quantified using two different field based methods. The DMI values obtained for the material stack W/Co$_{20}$Fe$_{60}$B$_{20}$/MgO from both methods are in good agreement and the sizeable DMI stabilizes skyrmions at room temperature. Using W/Co$_{40}$Fe$_{40}$B$_{20}$/MgO a strongly reduced DMI is found. We finally show that metastable skyrmions can be generated in this material at room temperature, and can be moved by current pulses due to spin orbit torques.

This work has been funded by the European Community under the Marie-Curie Seventh Framework program – ITN "WALL" (Grant No. 608031), the DFG (in particular SFB TRR173 Spin+X) and the Graduate School of Excellence Materials Science in Mainz (Mainz). The authors would like to thank R. Khan and K. Shahbazi for their helpful discussions.